\documentclass[openacc]{rstransa}
\usepackage{lmodern}
\usepackage{booktabs}
\usepackage{pdflscape}   
\usepackage{afterpage}  
\usepackage{url}

\newcommand{\kms}{km\,s$^{-1}$}

\titlehead{Research}

\begin{document}

\title{Decay Length of Slow Magnetoacoustic Waves in the Solar Corona using SDO/AIA and SolO/EUI}

\author{
R. L. Meadowcroft$^{1}$, 
N. Narang$^{2}$, 
D. Berghmans$^{2}$,
M. Dominique$^{2}$
and V. M. Nakariakov$^{1, 3}$}

\address{
$^{1}$CFSA, Physics Department, University of Warwick, Coventry CV4 7AL, UK\\
$^{2}$STCE -- SIDC, Royal Observatory of Belgium, Ringlaan -3- Av. Circulaire, Brussels, 1180, Belgium\\
$^{3}$G-LAMP NEXUS Institute/School of Space Research, Kyung Hee University,  Yongin, 17104, Republic of Korea
}

\subject{Astrophysics, Solar system, Plasma physics, Wave motion}


\corres{V. M. Nakariakov\\
\email{V.Nakariakov@warwick.ac.uk}}

\begin{abstract}
Slow magnetoacoustic waves have been observed in the solar corona for decades; yet, existing theories do not fully explain the wide range of observed decay lengths, and measurements from different instruments frequently disagree.
We present two events with simultaneous SDO/AIA and Solar Orbiter/EUI-HRIEUV observations of slow waves in sunspot-anchored coronal fan feathers. In Event~1, the LoS are separated by $16.2^\circ$, while in Event~2, they are almost parallel.
Using time–distance analysis, we identify waves with periods in the 3-minute band and quantify wave decay using the exponential folding length.
When LoS are non-parallel, the ratio of decay lengths measured by the two instruments, $\lambda_\mathrm{ratio} = \lambda_{\mathrm{AIA}} / \lambda_{\mathrm{EUI}}$ varies widely, with both $\lambda_\mathrm{ratio} > 1$ and $\lambda_\mathrm{ratio} < 1$ observed. 
However, when the LoS are almost parallel, the decay lengths agree with one another, and $\lambda_\mathrm{ratio} \approx 1$. 
These results show that the angle between the LoS and the wave propagation direction is key to determining the apparent decay of slow waves, a factor that has not yet been systematically accounted for. Since statistical studies of decay lengths constrain coronal energy-loss mechanisms, observational effects beyond simple projection must be corrected to reliably infer the underlying physical dissipation processes.
\end{abstract}


\begin{fmtext}
\end{fmtext}

\maketitle

\section{Introduction}
\label{sec:intro}

A common dynamic feature of the solar corona is propagating EUV emission intensity disturbances. This wave phenomenon has been routinely observed, most frequently in bright features of coronal fans (i.e. feathers) and in loops anchored in sunspots (e.g. \cite{1999SoPh..186..207B, 2000A&A...355L..23D, 2001A&A...370..591R, 2013ApJ...778...26U
}), as well as in polar plumes (e.g. \cite{1998ApJ...501L.217D, 2011A&A...528L...4K}).
The observed properties of these waves, namely, that the direction of propagation coincides with the apparent direction of the local magnetic field, and that the instantaneous phase speed is of the order of several tens of \kms and appears to be subsonic, indicate the slow magnetoacoustic nature of this wave phenomenon. The oscillation periods range from a few to several tens of minutes (e.g., \cite{2021SSRv..217...76B}). The increasing and decreasing slopes of the EUV emission intensity are symmetric, i.e., the perturbation is almost harmonic. Often, the wave patterns consist of a number of oscillation cycles, up to several tens, without a detected period variation (e.g., \cite{2025MNRAS.536.3192M}). The subsonic propagating disturbances in EUV emission intensity should not be confused with supersonic propagating disturbances, or quasi-periodic fast propagating (QFP) waves, which typically exhibit shorter periods and phase speeds of several hundred \kms (see, e.g., \cite{2022SoPh..297...20S} for a review). Coronal slow waves appear also in a standing and/or sloshing form, e.g., \cite{2019ApJ...874L...1N, 2021SSRv..217...34W} for reviews. 

The interpretation of subsonic propagating disturbances in EUV emission intensity in terms of slow magnetoacoustic waves is based on the developed theory of magnetohydrodynamic (MHD) wave propagation along a field-aligned plasma non-uniformity, i.e. a transverse structuring stretched along the magnetic field (see, e.g., \cite{1983SoPh...88..179E}).
In the optically thin radiation regime typical of the solar corona, the density perturbations intrinsic to the slow mode readily produce EUV emission intensity variations. The phase speed of slow waves guided by a field-aligned plasma non-uniformity, such as a feather or a plume, is subsonic.
In the long-wavelength limit, when the parallel wavelength is much longer than the transverse width of the non-uniformity, the phase speed approaches the tube (or cusp) speed. This speed is determined by the local sound and Alfv\'en speeds. In the low-$\beta$ plasma typical of the corona, the tube speed almost coincides with the sound speed. In this so-called infinite field approximation, coronal slow waves can be adequately modelled in terms of 1D acoustics, see, e.g., \cite{2000A&A...362.1151N, 2004A&A...415..705D}. 
The departure of the observed phase speed from the sound speed, inferred from the temperature of the emitting plasma, is attributed to the angle between the wave vector, which is aligned with the guiding plasma non-uniformity and the LoS. Furthermore, the local sound speed can also be modified by the effect of thermal misbalance and therefore may differ from its adiabatic value \cite{2019PhPl...26h2113Z}. More advanced modelling accounts for the coupling between parallel slow and entropy waves \cite{2021SoPh..296...96Z}. 

The narrowband nature of propagating slow magnetoacoustic waves is likely associated with the leakage of chromospheric oscillations along the magnetic field into the corona \cite{2011ApJ...728...84B}. In this scenario, the oscillation period is determined by the vertical non-uniformity of the plasma in the lower layers of the solar atmosphere (e.g., \cite{2023LRSP...20....1J}). The difference in the specific oscillation periods detected in individual coronal magnetic flux tubes anchored at different locations in the chromosphere further supports this interpretation \cite{2025MNRAS.536.3192M}.

Coronal propagating slow waves emanating from the chromosphere along magnetic flux tubes are subject to enhanced damping. In coronal active regions, the waves are no longer detected above a few tens of Mm (e.g. \cite{2014ApJ...789..118K}). Several dedicated studies have investigated mechanisms of slow-wave damping with thermal conduction identified as the dominant dissipation mechanism \cite{2004A&A...415..705D}. Thermal conduction acts to damp slow waves by transporting heat away from compressed, hotter regions to rarefied, cooler regions, smoothing temperature fluctuations induced by the wave. Compressive viscosity and optically thin radiation have a relatively minor contribution compared to thermal conduction \cite{2012SoPh..280..137A, 2021SoPh..296...20P}.

However, thermal conduction predicts decay lengths of the order of hundreds of Mm, which are significantly longer than those typically observed. Other works highlight alternative dominant damping mechanisms, such as the divergence of the waveguide with height \cite{2011ApJ...734...81M}. The dependence of the damping length on temperature also indicates that thermal conduction cannot be the dominant damping mechanism \cite{2019FrASS...6...57S}, suggesting that additional physical processes may play an important role in understanding slow-wave damping. For example, thermal misbalance results in wave damping due to the back reaction of the wave-induced misbalance between heating and cooling processes in the corona \cite{2017ApJ...849...62N}. It has been shown that, in certain regimes, thermal misbalance can yield damping rates that match observational values \cite{2019A&A...628A.133K}.

In addition to physical mechanisms that describe the dissipation of energy from the wave, observational effects could explain shorter-than-expected decay lengths. Phase mixing can occur when waves along adjacent field lines propagate at different local phase speeds, leading to progressive phase mismatch. In optically thin plasma, this results in an apparent decay in the amplitude when integrating along the LoS \cite{2005A&A...437L..47V}. Such effects highlight the challenge of distinguishing genuine physical damping from apparent, geometry-driven amplitude decay.

The first simultaneous detection of propagating slow waves with SDO/AIA and SolO/EUI-HRIEUV revealed that decay lengths differ significantly between the two instruments \cite{2024MNRAS.527.5302M}. In contrast, simultaneous observations with SDO/AIA and Hi-C showed consistent decay lengths across both instruments \cite{2025ApJ...987...57T}. Both studies analysed a single feather within a coronal fan, but differed in their viewing geometry: SDO/AIA and SolO/EUI-HRIEUV observed from separated vantage points, while SDO/AIA and Hi-C had nearly parallel lines of sight. These differences in both viewing geometry and instrument characteristics complicate direct comparison between the two studies.

Building on these previous findings, this study investigates the decay length of slow waves in several coronal fan feathers for two distinct events with different viewing geometries. Using this approach, we explore how viewing geometry affects apparent damping, thereby isolating geometric effects from instrument-related ones. 

The paper is structured as follows: in Section \ref{sec:obs}, we describe the observational information. Section \ref{sec:analysis} details the data analysis and results obtained. We then discuss these results in Section \ref{sec:discuss} and make our concluding remarks in Section \ref{sec:conc}.

\section{Observations and Data Processing}
\label{sec:obs}

\begin{figure}
    \centering
    \includegraphics[width=0.6\linewidth]{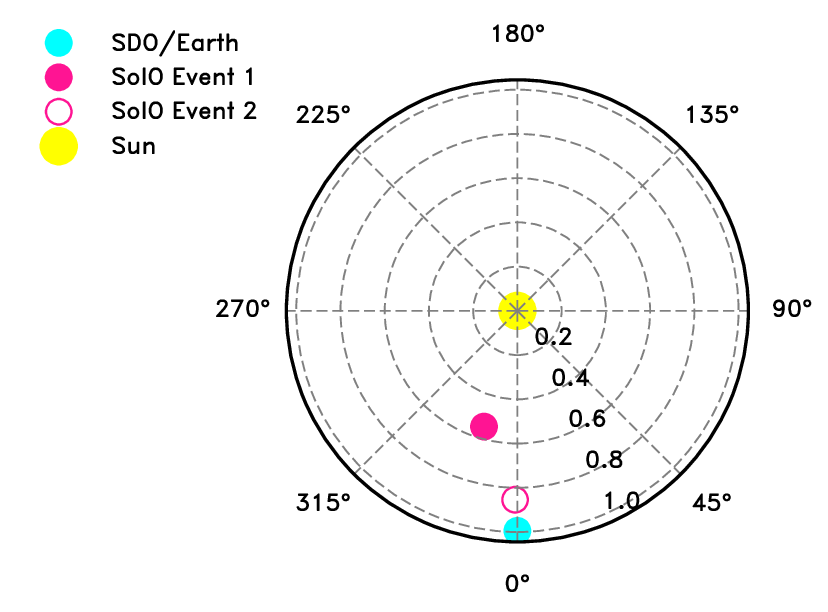}
    \caption{Heliocentric positions of the Sun (yellow), SDO/Earth (cyan), and SolO during Event~1 (filled magenta) and Event~2 (hollow magenta), shown in the ecliptic plane. Radial distances are in astronomical units. For Event~1, SolO and SDO are separated by $16.2^\circ$. In Event~2, the spacecraft have almost parallel LoS. Exact spacecraft positions are listed in Table~\ref{tab:spacecraft_positions}.}
    \label{fig:Spacecraft Position}
\end{figure}

\begin{table}
    \centering
    \begin{tabular}{lcccc}
        \toprule
        & \multicolumn{2}{c}{Event~1} & \multicolumn{2}{c}{Event~2} \\
        \cmidrule(lr){2-3} \cmidrule(lr){4-5}
        & AIA & HRIEUV & AIA & HRIEUV \\
        \midrule
        Observation Date & \multicolumn{2}{c}{October 24 2024} & \multicolumn{2}{c}{November 5 2021} \\
        Start time UT & \multicolumn{2}{c}{03:50:09} & \multicolumn{2}{c}{05:30:00} \\
        Duration (s) & \multicolumn{2}{c}{3600} & \multicolumn{2}{c}{3600} \\
        \midrule
        Cadence (s) & 12 & 3 & 12 & 5 \\
        Pixel Size (Mm\,pixel$^{-1}$) & 0.433 & 0.194 & 0.431 & 0.304 \\
        \midrule
        Longitude (deg) & 0.0 & -16.1 & 0.0 & -0.7 \\
        Latitude (deg)  & 5.1 & 8.0 & 3.9 & 2.0 \\
        Radius (AU)     & 0.995 & 0.545 & 0.992 & 0.854 \\
        LoS separation (deg) & \multicolumn{2}{c}{16.2} & \multicolumn{2}{c}{2.0} \\
        \bottomrule
    \end{tabular}
    \caption{Observational parameters and heliocentric spacecraft positions for the two events, shown in Fig.~\ref{fig:Spacecraft Position}. Longitudes, latitudes and radii are expressed in Stonyhurst heliocentric coordinates. The LoS separation represents the angular separation between the heliocentric position vectors of the two spacecraft.}
    \label{tab:spacecraft_positions}
\end{table}

\begin{figure*}
    \centering
    \includegraphics[width=0.9\linewidth]{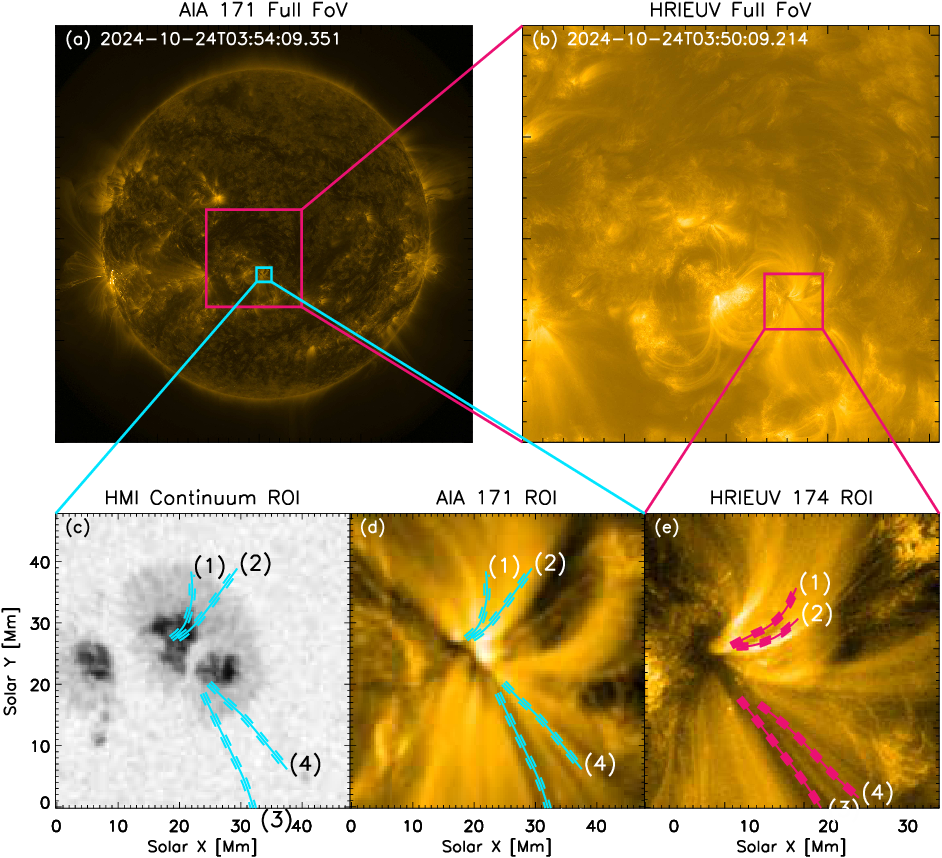}
    \caption{Event~1: \textbf{(a)} Full AIA 171\,\AA\, field of view (FoV) at the start of the observation (03:54:09~UT, 2024 October 24), corrected for light-travel time to correspond to the same emission time at the Sun. Cyan and magenta boxes indicate the region of interest (ROI) in AIA and HRIEUV, corresponding to panels (c, d) and (e), respectively. \textbf{(b)} Full HRIEUV 174\,\AA\, FoV at 03:50:09~UT. \textbf{(c)} HMI continuum at the same time as panel (a), showing that the active region of interest is rooted in several sunspots. \textbf{(d)} ROI in AIA 171\,\AA. \textbf{(e)} ROI in HRIEUV, co-aligned with AIA. The solid curves indicate the slits chosen for time--distance analysis; dashed curves show the additional pixels used. The slit width is 3\,pixels for AIA and 7\,pixels for HRIEUV to account for the difference in spatial resolution.}
    \label{fig:Setting the Scene}
\end{figure*}

\begin{figure*}
    \centering
    \includegraphics[width=0.9\linewidth]{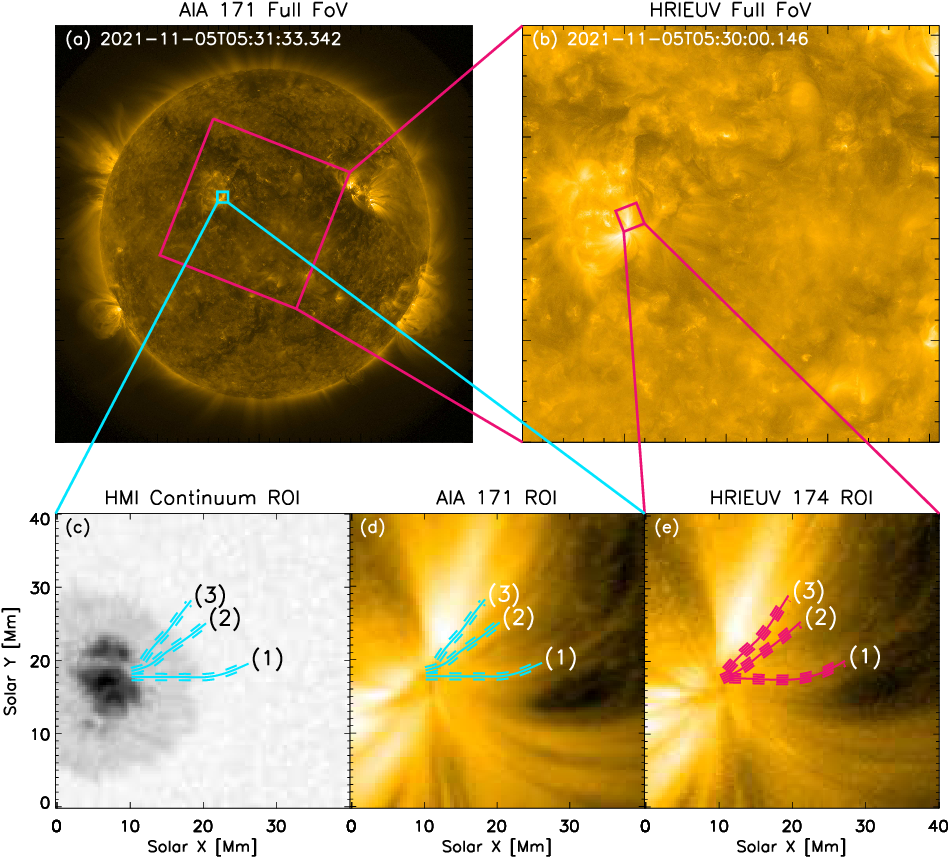}
    \caption{Event~2: As in Fig.~\ref{fig:Setting the Scene} but for 2021 November 5. The start time of the observation is given in panels (a) and (b), corrected for light travel time, corresponding to observing the same emission time at the Sun. In this event, AIA and HRIEUV have almost parallel LoS, and the slit width is 3\,pixels for AIA and 5\,pixels for HRIEUV to account for the difference in spatial resolution.}
    \label{fig:Setting the Scene Parallel}
\end{figure*}

Slow magnetoacoustic waves are observed propagating along several loop-like structures (``feathers'') of two coronal fans.
These waves were simultaneously observed by the \textit{Solar Dynamics Observatory}'s Atmospheric Imaging Assembly (SDO/AIA; \cite{2012SoPh..275...17L} ) and the \textit{Solar Orbiter}'s Extreme Ultraviolet Imager, High Resolution Imager in Extreme Ultraviolet (SolO/HRIEUV; \cite{2020A&A...642A...8R, euidatarelease6}). We make use of the AIA 171\,\AA\ and HRIEUV 174 \AA\,, and HRIEUV open filter channels, where the coronal fan and associated waves are observed with the highest contrast. The waves are observed as ripples along coronal feathers in extreme ultraviolet imaging data, and seen most clearly where the feathers appear to be anchored, emerge, and diverge.

SDO is in a geosynchronous orbit around Earth, and SolO follows an elliptical orbit around the Sun, allowing simultaneous observations at different angular separations between the two spacecraft. The positions of the spacecraft relative to the Sun in the ecliptic plane are illustrated in Fig.~\ref{fig:Spacecraft Position}. The position of SDO is shown in cyan, and the HRIEUV positions are shown by the solid and hollow magenta circles for Event~1 and Event~2, respectively. Table~\ref{tab:spacecraft_positions} lists the observational parameters and Stonyhurst heliocentric coordinates of the spacecraft for the two events. The Stonyhurst heliocentric coordinate system provides longitudes and latitudes that are measured relative to the Sun-Earth line in the ecliptic plane at the time of observation. These two events were selected because they correspond to non-parallel and almost parallel viewing angles, and are sufficiently long to reliably analyse wave properties. This allows us to investigate the effect of LoS geometry.

Event~1 focuses on the $\beta$-class active region (AR)\,13863 on October 24th 2024. HRIEUV was located at a heliocentric distance of 0.545\,AU, with a longitudinal position of $-16.1^{\circ}$ and a latitude of $8.0^{\circ}$. This corresponds to an angular separation of $16.2^{\circ}$ between the SDO and SolO LoS to the Sun. The HRIEUV observations were taken as part of the R\_SMALL\_MRES\_HCAD\_Sunspot-Oscillations Solar Orbiter Observing Plan (SOOP), and have a temporal resolution of 3\,s, four times higher than AIA's 12\,s cadence, and a pixel scale of 0.194\,Mm\,pixel$^{-1}$, providing approximately twice the spatial resolution of AIA, 0.433\,Mm\,pixel$^{-1}$. The observation lasted 1\,hour, limited by the length of the HRIEUV observing sequence, and began at 03:50:09\,UT.

Event~2 corresponds to the observation of $\alpha$-class AR\,12893 on 5th November 2021. In this event, AIA and HRIEUV have almost parallel LoS and have a separation angle of only $2.0^{\circ}$ (see Table~\ref{tab:spacecraft_positions} for exact spacecraft positions). This serves as a control case as it allows us to analyse the waves with two different instruments without the complication of significant projection effects. This event was earlier in SolO's lifetime, and hence the spacecraft was much closer to the Earth (0.854\,AU) and also at a smaller latitude ($2.0^{\circ}$). This observation does not correspond to a specific SOOP. The cadences are 5\,s with HRIEUV and 12\,s with AIA, and the pixel sizes are 0.431 and 0.304\,Mm\,pixel$^{-1}$ for AIA and HRIEUV, respectively, due to the close positioning of the two spacecraft. The observation lasts approximately 1\,hour starting at 05:30:00\,UT and the event length is again limited by the HRIEUV sequence length. 

The first frames of the AIA and HRIEUV observations are shown in Fig.~\ref{fig:Setting the Scene} (Event~1) and Fig.~\ref{fig:Setting the Scene Parallel} (Event~2). Panels (a) and (b) show the full AIA and HRIEUV field of view (FoV). The region of interest (ROI), as seen in the HMI continuum (c) and AIA 171\,\AA\ (d), is highlighted with a cyan box, while the corresponding HRIEUV 174\,\AA\ ROI (e) is shown in magenta. The cyan and magenta solid and dashed curves mark the slits used for time--distance analysis, described in Section~\ref{sec:analysis}.

The AIA and HRIEUV data require distinct processing steps. 
For AIA, level 1 files were downloaded from JSOC\footnote{\url{http://jsoc.stanford.edu/}} and converted to level 1.5 using the \texttt{aia\_prep} routine from SolarSoft. This process updates pointing information, removes the instrument roll angle, centres the image, and normalises the pixel size. Solar rotation was then corrected using \texttt{sdo\_track\_object}. The HRIEUV requires an additional processing step to correct for spacecraft jitters. Level 2 HRIEUV files were obtained using \texttt{sunpy-soar}\footnote{\url{https://www.astro.oma.be/doi/ROB-SIDC-SolO_EUI-DataRelease6.0_2023-01.html}}. The Level 2 images are calibrated and corrected for dark current, flat-fielding, and low-gain data is used to fill saturated pixels in the high-gain image and improve pointing instabilities.
The spacecraft jitter is still present in the Level 2 images and was corrected using \texttt{fg\_rigidalign} \cite{2022A&A...667A.166C}. The \texttt{fg\_rigidalign} procedure was applied twice: the first removed major drifts in the observation, and a second application of \texttt{fg\_rigidalign} removed any remnants of jitter and yielded a final, stable dataset. 

Several steps are also taken to align the data in time and space. To align in time, we must account for the light travel time difference between HRIEUV and AIA. This time difference is defined by $\Delta t = \Delta d / c$, where $\Delta d = d_{\mathrm{AIA}}-d_{\mathrm{EUI}}$ is the difference in distance from the Sun and AIA and EUI, and c is the speed of light. Using the EUI times as the reference, the number of AIA frames corresponding to $\Delta t$ is removed from the start of the AIA sequence to ensure time alignment. 
The AIA and HRIEUV images are aligned in space by first rotating the HRIEUV data to match the AIA orientation, see the HRIEUV FoV on AIA FoV in Fig.~\ref{fig:Setting the Scene Parallel}. Then, for each dataset, the first frame was used to select five distinctive points. Using the AIA image as a reference, the mean of the five x and y coordinates is calculated to obtain an average shift to align the images.

\section{Data Analysis and Results}
\label{sec:analysis}
\begin{figure}
    \centering
    \includegraphics[width=0.6\linewidth]{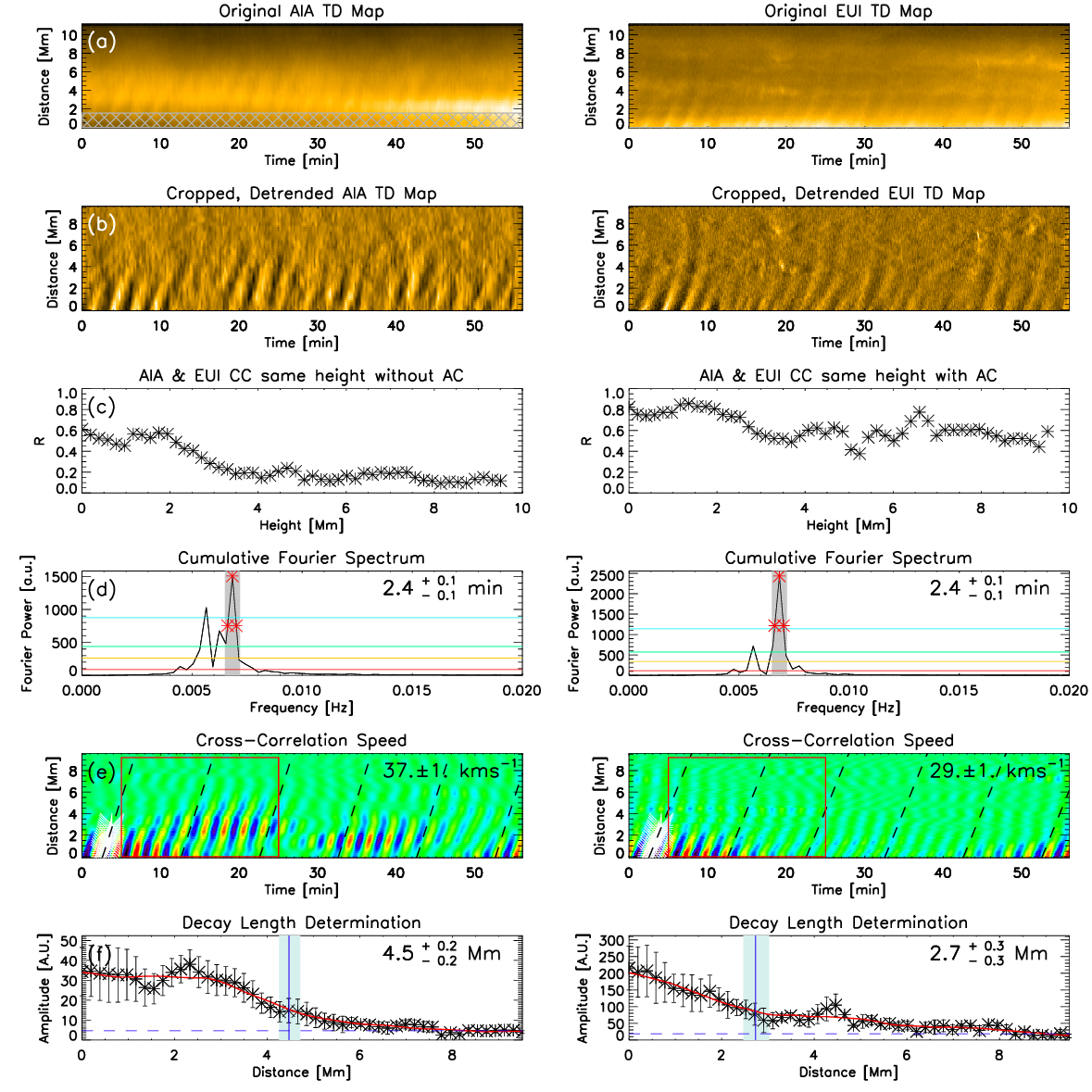}
    \caption{Wave property determination for Event~1, Slit 1 SDO/AIA (see Fig.~\ref{fig:Setting the Scene}). \textbf{(a)} Time--distance (TD) map interpolated to the HRIEUV resolution, the hashed region shows the area cropped to align AIA and HRIEUV TD maps. \textbf{(b)} Detrended TD map, obtained by subtracting a 250\,s running mean at each distance. \textbf{(c)} Cross-correlation coefficient (without auto-correlation) between AIA and HRIEUV data for each distance along the slit. \textbf{(d)} Cumulative Fourier spectrum from the detrended TD map. Red stars indicate the dominant period and its error (from FWHM). Horizontal red, amber, green, and blue lines show the 1, 3, 5, and 10\,$\sigma$ levels. \textbf{(e)} Filtered TD map, obtained by filtering the detrended TD map. The colour scheme is changed for visualisation. The white stars show the identified cross-correlation times and diagonal dashed lines illustrate the projected wave speed, determined by a linear fit to the cross-correlation times. \textbf{(f)} Average oscillation amplitude along slit for the region in the red box in (e). The red line shows the smoothed trend along the slit. Dashed horizontal line indicates the background intensity, the vertical blue line indicates the $e$-folding length, and the blue shaded region shows the corresponding error.}
    \label{fig:TD Analysis}
\end{figure}

\afterpage{%
\clearpage
\begin{landscape}

\begin{table}[p]
\centering
\renewcommand{\arraystretch}{1.1}

\begin{tabular}{lcccccccc}
    \toprule
        & \multicolumn{2}{c}{Slit 1} & \multicolumn{2}{c}{Slit 2} & \multicolumn{2}{c}{Slit 3} & \multicolumn{2}{c}{Slit 4} \\
        \cmidrule(lr){2-3} \cmidrule(lr){4-5} \cmidrule(lr){6-7} \cmidrule(lr){8-9}
         & AIA & EUI & AIA & EUI & AIA & EUI & AIA & EUI \\
        \midrule

        Period (min) 
            & $2.4^{+0.1}_{-0.1}$ & $2.4^{+0.1}_{-0.1}$ 
            & $2.4^{+0.1}_{-0.1}$ & $2.4^{+0.1}_{-0.1}$ 
            & $3.0^{+0.4}_{-0.1}$ & $3.0^{+0.1}_{-0.1}$ 
            & $2.4^{+0.1}_{-0.1}$ & $2.4^{+0.2}_{-0.1}$ \\

        Speed (\kms) 
            & $37 \pm 1$ & $34 \pm 2$
            & $32 \pm 0.5$ & $38 \pm 4$
            & $54 \pm 3$ & $66 \pm 4$
            & $45 \pm 5$ & $52 \pm 4$ \\

        Decay Length (Mm) 
            & $4.5 \pm 0.3$ & $2.6 \pm 0.3$
            & $3.8 \pm 0.6$ & $3.6 \pm 0.5$
            & $2.2 \pm 0.1$ & $5 \pm 1$
            & $1.5 \pm 0.1$ & $3.0 \pm 0.2$ \\

        \addlinespace[0.3cm]

        Speed Ratio (AIA/EUI)
            & \multicolumn{2}{c}{$1.1 \pm 0.1$} 
            & \multicolumn{2}{c}{$0.8 \pm 0.1$}
            & \multicolumn{2}{c}{$0.8 \pm 0.1$}
            & \multicolumn{2}{c}{$0.9 \pm 0.1$} \\

        Decay Length Ratio (AIA/EUI)
            & \multicolumn{2}{c}{$1.7 \pm 0.3$}
            & \multicolumn{2}{c}{$1.1 \pm 0.2$}
            & \multicolumn{2}{c}{$0.4 \pm 0.1$}
            & \multicolumn{2}{c}{$0.5 \pm 0.1$} \\

        Speed Ratio / Decay Length Ratio
            & \multicolumn{2}{c}{$0.6 \pm 0.1$}
            & \multicolumn{2}{c}{$0.8 \pm 0.2$}
            & \multicolumn{2}{c}{$1.8 \pm 0.5$}
            & \multicolumn{2}{c}{$1.8 \pm 0.3$} \\

        \bottomrule
\end{tabular}
\caption{Determined wave parameters for Event~1, Slits 1--4, as highlighted in Fig.~\ref{fig:Setting the Scene}. For each slit, the oscillation period, propagation speed, and decay length are determined separately for the AIA and HRIEUV datasets. The period is determined using the cumulative Fourier method, the speed is calculated using cross-correlation, and the decay length is defined as the distance over which the maximum amplitude decreases by a factor of $e$. The bottom three rows show the ratios between the two instruments (AIA/HRIEUV): the speed ratio, decay length ratio, and the ratio of these two quantities (speed ratio/decay length ratio). These ratios quantify the relative influence of projection effects on each instrument's observations.}
\label{tab:waveproperties}
\end{table}

\begin{table}[p]
\centering
\begin{tabular}{lcccccc}
        \toprule
        & \multicolumn{2}{c}{Slit 1} & \multicolumn{2}{c}{Slit 2} & \multicolumn{2}{c}{Slit 3} \\
        \cmidrule(lr){2-3} \cmidrule(lr){4-5} \cmidrule(lr){6-7} 
         & AIA & EUI & AIA & EUI & AIA & EUI \\
        \midrule

        Period (min) 
            & $2.5^{+0.1}_{-0.1}$ & $2.5^{+0.1}_{-0.4}$ 
            & $2.5^{+0.1}_{-0.1}$ & $2.5^{+0.1}_{-0.1}$ 
            & $3.0^{+0.2}_{-0.3}$ & $3.0^{+0.1}_{-0.1}$ \\

        Speed (\kms) 
            & $43 \pm 2$ & $37 \pm 2$
            & $37 \pm 4$ & $40 \pm 2$
            & $25 \pm 1$ & $22 \pm 5$ \\

        Decay Length (Mm) 
            & $2.1 \pm 0.1$ & $2.1 \pm 0.1$
            & $2.1 \pm 0.1$ & $1.8 \pm 0.1$
            & $1.7 \pm 0.1$ & $1.5 \pm 0.1$ \\

        \addlinespace[0.3cm]

        Speed Ratio (AIA/EUI)
            & \multicolumn{2}{c}{$1.2 \pm 0.1$} 
            & \multicolumn{2}{c}{$0.9 \pm 0.1$}
            & \multicolumn{2}{c}{$1.1 \pm 0.3$} \\

        Decay Length Ratio (AIA/EUI)
            & \multicolumn{2}{c}{$1.0 \pm 0.1$}
            & \multicolumn{2}{c}{$1.2 \pm 0.1$}
            & \multicolumn{2}{c}{$1.1 \pm 0.1$} \\

        Speed Ratio / Decay Length Ratio
            & \multicolumn{2}{c}{$1.1 \pm 0.1$}
            & \multicolumn{2}{c}{$0.8 \pm 0.1$}
            & \multicolumn{2}{c}{$1.0 \pm 0.3$} \\

        \bottomrule
\end{tabular}
\caption{As in Table~\ref{tab:waveproperties} but for Event~2, Slits 1--3, where the AIA and HRIEUV LoS are almost parallel.}
\label{tab:waveproperties_parallel}
\end{table}

\end{landscape}
\clearpage
}%

\subsection{Selecting slit positions}
To compare slow wave properties measured with AIA and HRIEUV, we must ensure we are observing the same wave with each instrument. This is made easier with the use of time--distance (TD) maps, produced by selecting slits parallel to the direction of wave propagation. Due to the projection effects of 3D structures, the same geometric slit placement on both data sets does not guarantee the same structure is observed. To combat this, the slit positions are selected manually by following the feather direction and using a correlation-based approach and slit match criteria.

Additionally, the width of the slits used for TD analysis differ between the AIA and EUI datasets to account for the differences in spatial resolutions. The slit widths are chosen to be an odd number of pixels to ensure symmetry around the central slit position. For Event~1, the ratio of the AIA to EUI pixel sizes is 2.22, which is best matched by a slit width of 3\,pixels for AIA and 7\,pixels for EUI. For Event~2, the ratio is 1.42 and a slit width of 3\,pixels in AIA and 5\,pixels for EUI is used.

First, the AIA TD maps were interpolated to the corresponding HRIEUV spatial and temporal resolutions for Event~1 and Event~2 using the IDL function \texttt{interpol}. These interpolated TD maps are used for the subsequent analysis for the wave period, speed and decay length.

To align the TD maps at 0 Mm and remove spatial offsets, the cross-correlation was computed by fixing the base of one dataset and shifting the other by 1 pixel along the TD map, then repeating with the other base fixed. A peak at a non-zero offset indicated a mismatch, and the TD map was cropped accordingly (see the grey hashed region in Fig.~\ref{fig:TD Analysis}\,(a)). The process was then repeated using matched slit positions, shown in Fig.~\ref{fig:TD Analysis}\,(d). Additionally, the autocorrelation function for each dataset was computed to reduce the influence of transient features and noise, but as it can overestimate when the signal is small compared to the noise, it served only as a supplementary check.

A slit pair was defined as a `match’ if it met the following criteria: 
\begin{enumerate}
    \item Cross correlation without auto-correlation $> 0.6$,
    \item Cross correlation with auto-correlation $> 0.9$,
    \item Periods agree within uncertainties,
    \item Fourier spectra qualitatively similar.
\end{enumerate}
These slit match criteria are not always satisfied. Using this approach, we find four suitable feathers in Event~1 and three in Event~2 for further analysis.

\subsection{Wave Period}
The wave period is determined using the cumulative Fourier method \cite{2025MNRAS.536.3192M}. First, we remove long-term trends by subtracting a 250\,s running mean. This window removes long-period trends without affecting features of interest, isolating periods shorter than approximately 4\,min, see Fig.~\ref{fig:TD Analysis} (b). Next, we create a Fourier spectrum for each distance of the detrended TD map up to 2\,Mm. These spectra are summed to form the cumulative Fourier spectrum shown in Fig.~\ref{fig:TD Analysis} (d). We identify the dominant peak in the 0--0.02\,Hz range as the oscillation period. The uncertainty of the period is taken as the full width at half maximum (FWHM) of the peak; both the period and uncertainty are demonstrated as red stars in Fig.~\ref{fig:TD Analysis}\,(d). The red, amber, green, and blue lines indicate the 1, 3, 5, and 10\,$\sigma$ levels, where $\sigma$ is the standard deviation of the Fourier spectrum. For AIA Slit 1, the dominant period was $2.4\pm0.1$\,min, consistent with the typical range for slow magnetoacoustic waves in sunspot-associated coronal fans. Here, the error is symmetric, but in some cases it is not, due to the Fourier spectrum's limited resolution, which is determined by the data's length. We repeated this process for each slit. The period values for all slits are given in Table~\ref{tab:waveproperties} for Event~1 and in Table~\ref{tab:waveproperties_parallel} for Event~2.

\subsection{Projected Wave Speed}
The projected wave speed is determined using the cross-correlation method. This method uses the filtered TD map to highlight the dominant oscillation period, as shown in Fig.~\ref{fig:TD Analysis} (e). The filtered TD map is obtained by filtering the detrended TD map in frequency space using a Gaussian filter with a width equal to $f/10$ where $f$ is the dominant frequency determined in the cumulative Fourier spectrum. This width is demonstrated by the grey shaded region in Fig.~\ref{fig:TD Analysis} (d). The cross-correlation function was computed between the signal at 0\,Mm along the slit and the signal at each pixel, up to 5\,Mm. The times at which the cross-correlation function is maximum are fitted with a linear function, with the slope corresponding to the projected wave speed. The 1\,$\sigma$ error of the linear fit gives the error of the projected wave speed. In Fig.~\ref{fig:TD Analysis}, the white stars show the identified cross-correlation times for each distance, and the dashed black lines in panel (e) demonstrate the determined wave speed. For AIA Slit 1, the derived speed is $37 \pm 1$\,\kms; this speed is consistent with the sound speed in a coronal fan projected onto the plane of the sky. This process is repeated for each of the seven slits, and the speed values are given in Table~\ref{tab:waveproperties} for Event~1 and Table~\ref{tab:waveproperties_parallel} for Event~2.

\subsection{Decay Length}
The decay length is determined using the exponential folding length \cite{2024MNRAS.527.5302M}. 
We selected a small region from the filtered TD map where the oscillation amplitude was fairly stable in time to minimise errors when calculating the decay length. For Event~1 we used select 5—25\,min and 0—10\,min for Event~2.
This region is different for each event, as the time over which the decay length remained constant was shorter in Event~2. Then for each distance, we computed the mean amplitude, with its standard deviation taken as the uncertainty. The amplitudes (black stars) and their associated errors are shown in Fig.~\ref{fig:TD Analysis}\,(f). These points were then smoothed using a running mean with a 3\,Mm window to define the trend, shown by the red curve in panel (f).
The exponential folding length is the distance between the maximum amplitude of the smoothed trend, $A_{\mathrm{max}}$, and $A_{\mathrm{max}}/e$, the distance at which the amplitude has decreased by a factor of $e$. To extend the previous method, we estimated the background noise level as the mean amplitude over the final 2\,Mm of the TD map, assuming the wave had fully decayed at this distance. This background amplitude is subtracted from all amplitude values before calculating the decay length. 
The error in the decay length is determined by taking the mean of the errors of all measurements, $\Delta A$, and calculating the distance at which the smoothed amplitude drops below $(A_{\mathrm{max}}\pm \Delta A)/e$. 
This process is repeated for each of the seven slits, and the decay length values are given in Table~\ref{tab:waveproperties} for Event~1 and Table~\ref{tab:waveproperties_parallel} for Event~2.

We define the speed, decay-length, and speed/decay-length ratios as the AIA values divided by the corresponding HRIEUV values in Tables~\ref{tab:waveproperties} and \ref{tab:waveproperties_parallel}. The uncertainties were propagated assuming independent Gaussian errors using standard first-order error propagation. The implications of these ratios are discussed in Section\,\ref{sec:discuss}.

\section{Discussion}
\label{sec:discuss}
\begin{figure*}
    \centering
    \includegraphics[width=1\linewidth]{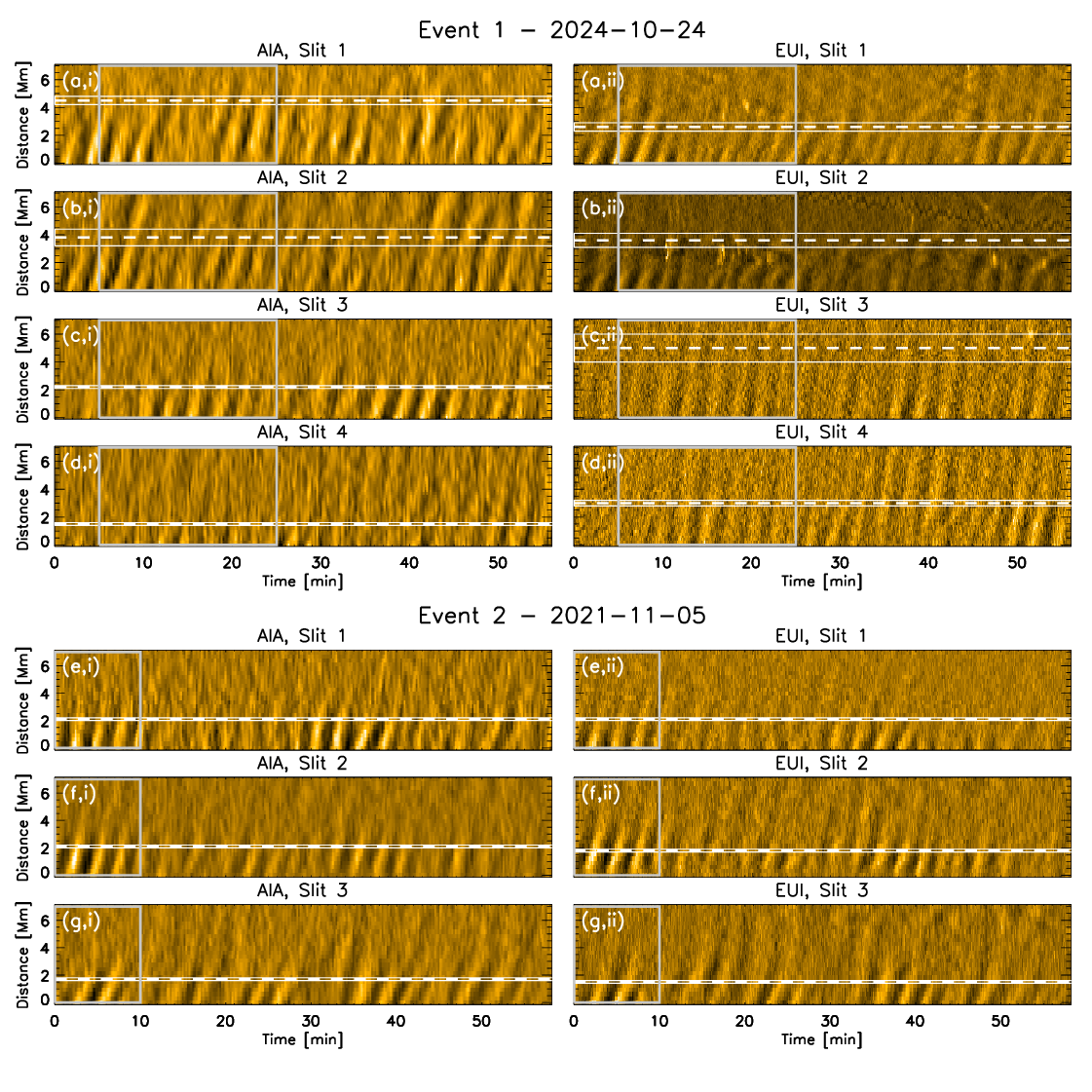}
    \caption{Detrended, aligned TD maps obtained for Event~1, Slits 1--4 (panels a to d), and Event~2, Slits 1--3 (panels e to g) as defined in Figs. \ref{fig:Setting the Scene} and \ref{fig:Setting the Scene Parallel}. The left panels (i) correspond to AIA data, interpolated to the same resolution as the HRIEUV data, and the right panels (ii) to the HRIEUV data. The dashed white lines show the measured slow wave decay length, the solid lines show the associated uncertainty range, and the grey boxes illustrate the regions used to calculate the decay length. All TD maps are cropped so that a distance of 0\,Mm corresponds to the maximum amplitude value.}
    \label{fig:decay_lengths}
\end{figure*}

\begin{figure}
    \centering
    \includegraphics[width=0.65\linewidth]{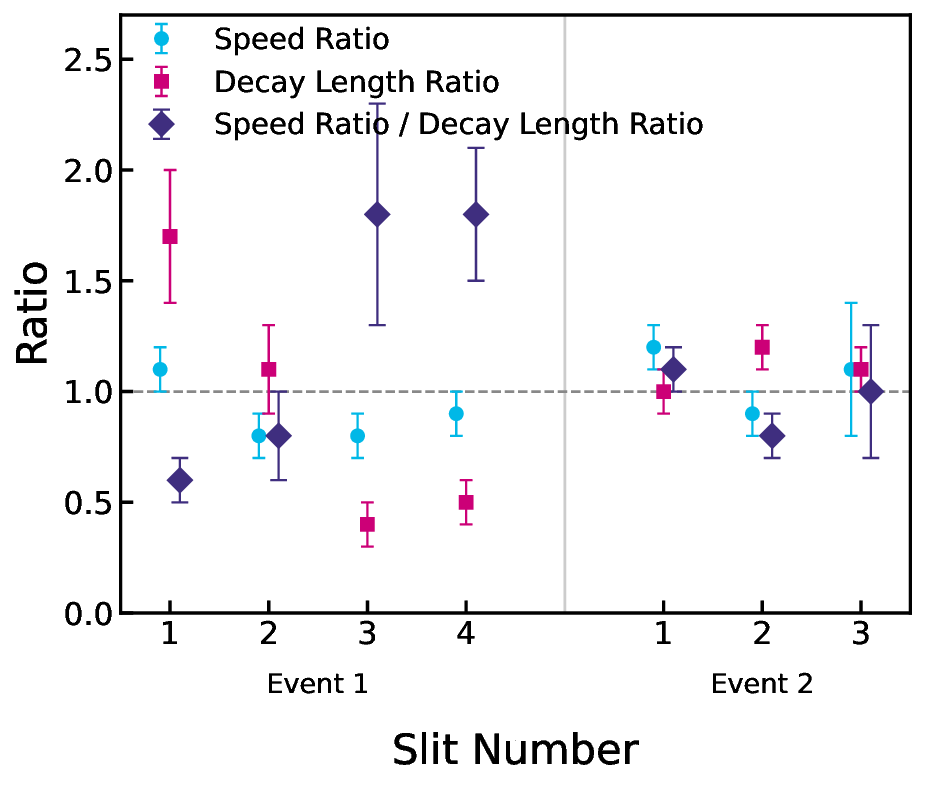}
    \caption{Ratios of wave speeds and decay lengths listed in Table~\ref{tab:waveproperties} (AIA/EUI) for Event~1, Slits 1--4 and Event~2, Slits 1--3. Blue circles correspond to the speed ratio, pink squares to the decay length ratio, and purple diamonds correspond to the ratio of these two quantities (speed ratio/decay length ratio). The horizontal dashed line indicates where the ratio equals unity, corresponding to agreement between the projected quantities. The vertical line separates Event~1 and Event~2.}
    \label{fig:ratios}
\end{figure}

In this study, we investigated the decay length of propagating slow waves in a coronal fan using coordinated SDO/AIA and SolO/EUI-HRIEUV observations with two different viewing geometries. The identified propagating intensity disturbances have projected speeds below the local sound speed, and oscillation periods around 3\,minutes. Both of which support the classification as slow magnetoacoustic waves, as seen in several past studies (see, e.g., \cite{2009SSRv..149...65D, 2016GMS...216..395W, 2021SSRv..217...76B} for reviews). We reveal an important, currently unexplained difference in the slow-wave decay length, observed only when the two instruments have non-parallel LoS. In the following discussion, we explore the challenges of comparing these datasets, the reliability of methods for measuring the decay length, and the extent to which projection effects can account for the observed variations.

\textbf{Difficulty identifying matching slits.} 
Although surface alignment of AIA and HRIEUV observations is possible using the pointing metadata in the image files, we do not perform such an alignment in this work. Even with precise surface-level disc alignment, the 3D nature of the coronal fan means that alignment of coronal loops and other features is not possible due to projection effects and integration along the LoS. 
This is a limiting factor in interpreting differences between AIA and HRIEUV observations. The process of identifying corresponding slit positions for TD analysis is a manual and time-consuming task. We assign `slit-match criteria' to ensure that the slow waves we are studying are matched in both datasets. This criterion limits the regions we can compare for slow waves, and we identify a total of 7 matched slits: 4 in Event~1 and 3 in Event~2. Although this limits the number of events we can study, it ensures confidence that the same wave is being observed.
The development of a tool or technique to rigorously identify the presence, position, and direction of slow magnetoacoustic waves would be a great addition to the field of MHD seismology (e.g. \cite{2024RvMPP...8...19N}) and a potential avenue for future research. Furthermore, a Machine Learning method was presented by \cite{2025ApJS..281...12B} that can identify the presence or absence of kink oscillations; this could be extended to identify the presence, position, and direction of slow waves. 

\textbf{Method for determining decay length.} 
Once confident that we were observing the same waves with both instruments, we determined the wave parameters. The techniques for determining the wave's speed and period are widely used in the field. 
To determine the decay length, we identify the height at which the wave amplitude decreases by a factor of $e$, i.e., the $e$-folding length. 
Previous studies have typically employed two approaches to estimate the decay length: the phase-tracking method and the amplitude-tracking method as described in \cite{2019FrASS...6...57S}. The phase-tracking method involves fitting an exponentially decaying sinusoidal function to the spatial profile, averaged over a selected time interval. In contrast, the amplitude-tracking method estimates the wave amplitude at each spatial position, often using the standard deviation, and then fits an exponential decay to the resulting amplitude profile. 
However, these approaches assume that the amplitude decreases exponentially along the entire propagation path. In our observations, this assumption is not always valid. For example, in some cases the slow wave maintains an approximately constant amplitude over several Mm (e.g., Fig.~\ref{fig:TD Analysis}\,(f)) before any clear decay is observed. Fitting a single exponential function over the full distance would therefore be inappropriate, while restricting the fit to only the decaying portion would yield a decay length that does not represent the full spatial extent over which the wave is visible. Furthermore, the amplitude profile is not always well described by an exponential function, leading to poor fits and unreliable decay length estimates. For these reasons, we adopt the $e$-folding length as a more robust diagnostic, which characterises the distance over which the wave amplitude decreases significantly, without assuming a globally exponential behaviour.

\textbf{Comparing decay length with AIA and HRIEUV.} 
The TD maps for each slit in Events 1 and 2 are shown in Fig.~\ref{fig:decay_lengths}. For each slit, the decay length is shown by the white dashed horizontal line, and the region of the error is shown with the horizontal white solid lines. Immediately, it appears that there is more consistency between the decay length measurements in Event~2.
It is not possible to directly compare the measured decay lengths and speeds from the AIA and HRIEUV data, especially in Event~1, where the LoS of the two instruments are not parallel. This is because we measure the component of the speed or decay length in the plane of the sky, perpendicular to the LoS.
However, we can compare the ratio of the measured speeds and decay lengths to probe the effects of projection and viewing geometry. 
For each slit, we determine the ratio 
$$v_{\mathrm{ratio}} =  \frac{v_{\mathrm{AIA}}}{v_{\mathrm{EUI}}}= \frac{v\sin(\theta_{\mathrm{AIA}})}{v\sin(\theta_{\mathrm{EUI}})},$$
where $v$ is the absolute wave speed, $v_{\mathrm{AIA}}$ and $v_{\mathrm{EUI}}$ are the projected wave speeds, and $\theta_{\mathrm{AIA}}$ and $\theta_{\mathrm{EUI}}$ are the angles between the LoS and the wave vector with AIA and HRIEUV, respectively. The same logic can be applied to the decay length: 
$$\lambda_{\mathrm{ratio}} =  \frac{\lambda_{\mathrm{AIA}}}{\lambda_{\mathrm{EUI}}}= \frac{\lambda \sin(\theta_{\mathrm{AIA}})}{\lambda \sin(\theta_{\mathrm{EUI}})},$$
where $\lambda$ is the absolute decay length, and $\lambda_{\mathrm{AIA}}$ and $\lambda_{\mathrm{EUI}}$ are the projected decay lengths. 
For each feather, $\theta_{\mathrm{AIA}}$ and $\theta_{\mathrm{EUI}}$ differ, and therefore the quantities $v_{\mathrm{ratio}}$ and $\lambda_{\mathrm{ratio}}$ are not expected to be equal. However, for a given feather, taking the ratio $v_{\mathrm{ratio}}/\lambda_{\mathrm{ratio}}$ should remove the effect of projection, provided that projection is the only mechanism influencing the measured velocities and decay lengths. That is:
$$R =\frac{v_{\mathrm{ratio}}}{\lambda_{\mathrm{ratio}}}=1.$$
This is not always the case. The values of $v_{\mathrm{ratio}}$, $\lambda_{\mathrm{ratio}}$, and $R$ and given in Tables \ref{tab:waveproperties} and \ref{tab:waveproperties_parallel} and also plotted in Fig.~\ref{fig:ratios}. The blue circles, magenta squares, and purple diamonds show $v_{\mathrm{ratio}}$, $\lambda_{\mathrm{ratio}}$, and $R$, respectively. The horizontal dashed grey line indicates where the ratio is equal to 1 and the vertical line separates the data from Event~1 and Event~2. 
We see that in Event~1, where the LoS are separated by an angle of $16.2^{\circ}$, for Slit 1 and 2 $R<1$ and for Slit 3 and 4 $R>1$. This suggests that a factor other than projection is influencing the wave speed or the decay length, and that there are additional physical or observational effects we do not yet fully understand. 
Because the $\lambda_{\mathrm{ratio}}$ values vary much more than the $v_{\mathrm{ratio}}$ values, we suggest that the missing physics is in the decay length measurement and not the wave speed. 

\textbf{Speculations for decay length mismatch.}
A case was presented where the decay length measured with EUI was significantly larger than the decay length measured with AIA in \cite{2024MNRAS.527.5302M}. Using our definition of $R$ above and the values in their study gives $R\approx1.6$. However, because the work studied only a single feather of a coronal fan, it was not possible to conclude whether the discrepancy in the decay length arose from instrumental sensitivity, viewing orientation, or intrinsic loop structure. 
In contrast, our study examines two cases of several distinct feathers within two coronal fans, simultaneously observed with the same instrumental setup. 
Interestingly, the deviation of $R$ from 1 doesn't appear to be systematic. 
It does not appear that the decay length measured with HRIEUV is consistently larger than that measured with AIA, even after accounting for projection effects by taking the ratio of the projected speed and decay length. This suggests that differences in response functions or instrumental sensitivity alone cannot explain the observed behaviour.
For Event~2, where the LoS are almost parallel, we see much more agreement; the values of $R$ sit at or very close to 1. 
Because all four wave observations in Event~1 share the same instrumental configuration, the only remaining geometric factor that varies between them is the orientation of each feather relative to the line of sight (LoS). However, the optically thin nature of the plasma introduces an additional complication. Different viewing angles result in the AIA and HRIEUV observations sampling different structures along the LoS, including overlying and underlying emission. These contributions may influence the measured wave speeds and decay lengths. This emphasises the importance of forward modelling, where the background plasma conditions can be controlled.

A recent work explored the decay length of propagating slow waves in a coronal fan using AIA and the \textit{High-resolution Coronal Imager} (Hi-C; \cite{2019SoPh..294..174R}) \cite{2025ApJ...987...57T}. They found that the decay lengths measured with each instrument were in agreement. In that observation, AIA and Hi-C viewed the structure from almost parallel LoS, and each instrument would experience the same projection effects. Our work is in agreement with this result. When AIA and HRIEUV are aligned (Event~2), we observe agreement in the decay-length measurements. 
While the present study exploits observations from instruments with non-parallel lines of sight, it is limited to two viewpoints and only two events. As demonstrated here, projection effects can significantly influence the measured decay lengths, and these effects vary depending on the orientation of individual structures within the same coronal fan. However, with only two viewing angles, it is not possible to fully disentangle projection effects from line-of-sight integration and intrinsic physical variations in the wave properties.

Future observations combining multiple instruments with different viewing geometries, such as STEREO-A/EUVI, Solar Orbiter/HRIEUV, and SDO/AIA, would enable a more systematic investigation of these effects. In particular, a larger sample of events observed over a range of viewing angles would allow the dependence of the measured wave speed and decay length on viewing geometry to be quantified, and help determine whether discrepancies between instruments arise from geometric effects or intrinsic physical differences (e.g., due to phase mixing effects or background plasma structuring).

Simultaneous observations from three distinct viewpoints would provide additional constraints on the three-dimensional geometry of the coronal structures. While two viewpoints can, in principle, be used to estimate the direction of propagation under certain assumptions, they do not uniquely determine the three-dimensional structure of the waveguide. Additionally, these results highlight that, without a comprehensive understanding of observational effects, it remains challenging to robustly determine the true propagation speed and decay length of slow magnetoacoustic waves from imaging data alone.

A numerical model showed that apparent damping of slow waves can arise from multi-thermal plasma within a plasma slab \cite{2024SoPh..299....2F}. An analytical work by \cite{2024A&A...683A.109V} also showed that a loop with a Gaussian temperature distribution can result in an apparent damping of slow waves, and this matched complementary numerical simulations. In this scenario, the different temperatures result in different sound speeds, causing an initially aligned wavefront to distort. This process decreases the apparent decay length because waves in neighbouring temperature plasmas become out of phase. Additionally, forward modelling by \cite{2024SoPh..299....2F} showed that observing the slab at different angles relative to its axis yields different measured decay lengths.
Here, we explore whether such phase mixing could be responsible for the mismatch in decay length we observe. Assuming there are two waves travelling at two extrema of temperatures within a coronal loop, 1\,MK and 0.6\,MK, and that the wave speed is given by $c_\mathrm{s}[kms^{-1}]=152\sqrt{T [\mathrm{MK}]}$. Consider a wave with oscillation period, $P = 3$\,mins, and wave number $k = 2\pi/Pc_\mathrm{s}$. Then, the distance at which two sinusoids of the form $\sin(kx-\omega t)$ first reach a phase difference of $\pi$ is given by:
$$\frac{\pi}{k_1-k_2}\approx 47\,\mathrm{Mm},$$
where $k_1$ and $k_2$ correspond to the wave number at 0.6\,MK and 1.0\,MK, respectively. This distance is substantially longer than the decay lengths observed in our fan. Hence, this effect alone does not explain fully the differences in decay length across slits with different viewing angles.

\textbf{Need for accurate decay length measurements.} 
Understanding coronal heating requires quantifying energy-loss mechanisms, such as thermal conduction and radiation. While statistical studies of slow wave decay lengths can constrain these mechanisms, observational effects beyond simple projection can artificially reduce the measured decay length. Without correcting for these effects, we cannot reliably infer the true energy loss processes governing coronal dynamics. While this work focuses on observational and geometric effects, intrinsic plasma processes such as thermal misbalance may also contribute to the apparent damping of slow waves, and should be explored in future modelling studies.

\section{Conclusions}
\label{sec:conc}

This study uses AIA and HRIEUV observations to investigate the properties of propagating slow magnetoacoustic waves in several feathers of coronal fans with non-parallel and almost parallel viewing geometries. We demonstrate that projection alone cannot account for the observed differences in projected decay lengths between the two instruments when observing with non-parallel LoS. Using the $e$-folding length as a diagnostic provides a more robust measure of amplitude decay than exponential fitting, including regions where the wave is not decaying or decay is non-exponential. 
These results highlight the importance of viewing angle and suggest that future coordinated observations, along with forward modelling and analytical models, are key to disentangling instrumental and physical effects.

\ack{
We thank the reviewers for their constructive comments and questions, which have helped to strengthen this work. We also thank Dmitrii Kolotkov for his helpful discussions in addressing one of the reviewer comments.
The data used in this work are courtesy of the NASA/SDO/AIA, and ESA \& NASA/Solar Orbiter/EUI. Solar Orbiter is a space mission of international collaboration between ESA and NASA, operated by ESA. The EUI instrument was built by CSL, IAS, MPS, MSSL/UCL, PMOD/WRC, ROB, LCF/IO with funding from the Belgian Federal Science Policy Office (BELSPO/PRODEX PEA 4000112292 and 4000134088); the Centre National d’Etudes Spatiales (CNES); the UK Space Agency (UKSA); the Bundesministerium für Wirtschaft und Energie (BMWi) through the Deutsches Zentrum für Luft- und Raumfahrt (DLR); and the Swiss Space Office (SSO).}

\dataccess{In this paper, we analysed data using the Interactive Data Language (IDL), SolarSoftWare (SSW, \cite{1998SoPh..182..497F}) package.
This research used version 7.0.1 of the SunPy open-source software package \cite{sunpy_community2020}, which was used to download EUI data via SOAR via Fido and \texttt{sunpy-soar}.
The EUI data are available at DOI: \url{https://doi.org/10.24414/z818-4163}. The SDO/AIA data are available at \url{http://jsoc.stanford.edu/}. }

\funding{R.L.M. and V.M.N. acknowledge funding from UK Research and Innovation under the UK government the Horizon Europe funding guarantee EP/Y037456/1 and the ERC grant 101201424 (ACDCSUN). R.L.M. also gratefully acknowledges the support of the Science and Technology Facilities Council (STFC) [grant number ST/X508871/1], and from the Royal Observatory of Belgium Guest Investigator Program funded by BELSPO in the framework of the ESA-PRODEX program, PEA SIDEX 4000145189. V.M.N. acknowledges support from the BK21 FOUR programme through the National Research Foundation of Korea (NRF) under the Ministry of Education (MoE) (Kyung Hee University, Human Education Team for the Next Generation of Space Exploration), and the Global-Learning \& Academic research institution for Master’s/PhD students, the Postdocs (G-LAMP) Program of the NRF grant funded by the MoE (RS-2025-25442355).}


\bibliographystyle{RS}
\bibliography{Decay_Length_Bib.bib}

@ARTICLE{2012SoPh..280..137A,
       author = {{Abedini}, A. and {Safari}, H. and {Nasiri}, S.},
        title = "{Slow-Mode Oscillations and Damping of Hot Solar Coronal Loops}",
      journal = {\solphys},
         year = 2012,
        month = sep,
       volume = {280},
       number = {1},
        pages = {137-151},
          doi = {10.1007/s11207-012-0054-1},
archivePrefix = {arXiv},
       eprint = {1206.0366},
 primaryClass = {astro-ph.SR},
       adsurl = {https://ui.adsabs.harvard.edu/abs/2012SoPh..280..137A}
}

@ARTICLE{2021SSRv..217...76B,
       author = {{Banerjee}, D. and {Krishna Prasad}, S. and {Pant}, V. and {McLaughlin}, J.~A. and {Antolin}, P. and {Magyar}, N. and {Ofman}, L. and {Tian}, H. and {Van Doorsselaere}, T. and {De Moortel}, I. and {Wang}, T.~J.},
        title = "{Magnetohydrodynamic Waves in Open Coronal Structures}",
      journal = {\ssr},
         year = 2021,
        month = oct,
       volume = {217},
       number = {7},
          eid = {76},
        pages = {76},
          doi = {10.1007/s11214-021-00849-0},
archivePrefix = {arXiv},
       eprint = {2012.08802},
 primaryClass = {astro-ph.SR},
       adsurl = {https://ui.adsabs.harvard.edu/abs/2021SSRv..217...76B}
}

@ARTICLE{2025ApJS..281...12B,
       author = {{Belov}, Sergey A. and {Zhong}, Yu and {Kolotkov}, Dmitrii Y. and {Nakariakov}, Valery M.},
        title = "{Detection of Kink Oscillations in Solar Coronal Loops by a Convolutional Neural Network}",
      journal = {\apjs},
         year = 2025,
        month = nov,
       volume = {281},
       number = {1},
          eid = {12},
        pages = {12},
          doi = {10.3847/1538-4365/ae0a34},
archivePrefix = {arXiv},
       eprint = {2509.15041},
 primaryClass = {astro-ph.SR},
       adsurl = {https://ui.adsabs.harvard.edu/abs/2025ApJS..281...12B}
}

@ARTICLE{1999SoPh..186..207B,
       author = {{Berghmans}, D. and {Clette}, F.},
        title = "{Active region EUV transient brightenings - First Results by EIT of SOHO JOP 80}",
      journal = {\solphys},
         year = 1999,
        month = may,
       volume = {186},
        pages = {207-229},
          doi = {10.1023/A:1005189508371},
       adsurl = {https://ui.adsabs.harvard.edu/abs/1999SoPh..186..207B}
}

@ARTICLE{2011ApJ...728...84B,
       author = {{Botha}, G.~J.~J. and {Arber}, T.~D. and {Nakariakov}, V.~M. and {Zhugzhda}, Y.~D.},
        title = "{Chromospheric Resonances above Sunspot Umbrae}",
      journal = {\apj},
         year = 2011,
        month = feb,
       volume = {728},
       number = {2},
          eid = {84},
        pages = {84},
          doi = {10.1088/0004-637X/728/2/84},
       adsurl = {https://ui.adsabs.harvard.edu/abs/2011ApJ...728...84B}
}

@ARTICLE{2022A&A...667A.166C,
       author = {{Chitta}, L.~P. and {Peter}, H. and {Parenti}, S. and {Berghmans}, D. and {Auch{\`e}re}, F. and {Solanki}, S.~K. and {Aznar Cuadrado}, R. and {Sch{\"u}hle}, U. and {Teriaca}, L. and {Mandal}, S. and {Barczynski}, K. and {Buchlin}, {\'E}. and {Harra}, L. and {Kraaikamp}, E. and {Long}, D.~M. and {Rodriguez}, L. and {Schwanitz}, C. and {Smith}, P.~J. and {Verbeeck}, C. and {Zhukov}, A.~N. and {Liu}, W. and {Cheung}, M.~C.~M.},
        title = "{Solar coronal heating from small-scale magnetic braids}",
      journal = {\aap},
         year = 2022,
        month = nov,
       volume = {667},
          eid = {A166},
        pages = {A166},
          doi = {10.1051/0004-6361/202244170},
archivePrefix = {arXiv},
       eprint = {2209.12203},
 primaryClass = {astro-ph.SR},
       adsurl = {https://ui.adsabs.harvard.edu/abs/2022A&A...667A.166C}
}

@ARTICLE{1998ApJ...501L.217D,
       author = {{DeForest}, C.~E. and {Gurman}, J.~B.},
        title = "{Observation of Quasi-periodic Compressive Waves in Solar Polar Plumes}",
      journal = {\apjl},
         year = 1998,
        month = jul,
       volume = {501},
       number = {2},
        pages = {L217-L220},
          doi = {10.1086/311460},
       adsurl = {https://ui.adsabs.harvard.edu/abs/1998ApJ...501L.217D}
}

@ARTICLE{2000A&A...355L..23D,
       author = {{De Moortel}, I. and {Ireland}, J. and {Walsh}, R.~W.},
        title = "{Observation of oscillations in coronal loops}",
      journal = {\aap},
         year = 2000,
        month = mar,
       volume = {355},
        pages = {L23-L26},
       adsurl = {https://ui.adsabs.harvard.edu/abs/2000A&A...355L..23D}
}

@ARTICLE{2009SSRv..149...65D,
       author = {{De Moortel}, I.},
        title = "{Longitudinal Waves in Coronal Loops}",
      journal = {\ssr},
         year = 2009,
        month = dec,
       volume = {149},
       number = {1-4},
        pages = {65-81},
          doi = {10.1007/s11214-009-9526-5},
       adsurl = {https://ui.adsabs.harvard.edu/abs/2009SSRv..149...65D}
}

@ARTICLE{2004A&A...415..705D,
       author = {{De Moortel}, I. and {Hood}, A.~W.},
        title = "{The damping of slow MHD waves in solar coronal magnetic fields. II. The effect of gravitational stratification and field line divergence}",
      journal = {\aap},
         year = 2004,
        month = feb,
       volume = {415},
        pages = {705-715},
          doi = {10.1051/0004-6361:20034233},
       adsurl = {https://ui.adsabs.harvard.edu/abs/2004A&A...415..705D}
}

@ARTICLE{1983SoPh...88..179E,
       author = {{Edwin}, P.~M. and {Roberts}, B.},
        title = "{Wave Propagation in a Magnetic Cylinder}",
      journal = {\solphys},
         year = 1983,
        month = oct,
       volume = {88},
       number = {1-2},
        pages = {179-191},
          doi = {10.1007/BF00196186},
       adsurl = {https://ui.adsabs.harvard.edu/abs/1983SoPh...88..179E}
}

@ARTICLE{2024SoPh..299....2F,
       author = {{Fedenev}, Viktor V. and {Nakariakov}, Valery M. and {Anfinogentov}, Sergey A.},
        title = "{Slow Magnetoacoustic Waves in Smoothly Nonuniform Coronal Plasma Structures}",
      journal = {\solphys},
         year = 2024,
        month = jan,
       volume = {299},
       number = {1},
          eid = {2},
        pages = {2},
          doi = {10.1007/s11207-023-02246-y},
       adsurl = {https://ui.adsabs.harvard.edu/abs/2024SoPh..299....2F}
}

@ARTICLE{1998SoPh..182..497F,
       author = {{Freeland}, S.~L. and {Handy}, B.~N.},
        title = "{Data Analysis with the SolarSoft System}",
      journal = {\solphys},
         year = 1998,
        month = oct,
       volume = {182},
       number = {2},
        pages = {497-500},
          doi = {10.1023/A:1005038224881},
       adsurl = {https://ui.adsabs.harvard.edu/abs/1998SoPh..182..497F}
}

@ARTICLE{2023LRSP...20....1J,
       author = {{Jess}, David B. and {Jafarzadeh}, Shahin and {Keys}, Peter H. and {Stangalini}, Marco and {Verth}, Gary and {Grant}, Samuel D.~T.},
        title = "{Waves in the lower solar atmosphere: the dawn of next-generation solar telescopes}",
      journal = {Living Reviews in Solar Physics},
         year = 2023,
        month = dec,
       volume = {20},
       number = {1},
          eid = {1},
        pages = {1},
          doi = {10.1007/s41116-022-00035-6},
archivePrefix = {arXiv},
       eprint = {2212.09788},
 primaryClass = {astro-ph.SR},
       adsurl = {https://ui.adsabs.harvard.edu/abs/2023LRSP...20....1J}
}

@misc{euidatarelease6,
  author       = {{Kraaikamp}, E. and {Gissot}, S. and  {Stegen}, K.  and {Mampaey}, B. and {Verbeeck}, F.  and  {Auch{\`e}re}, F. and {Berghmans}, D. },
  title        = {SolO/EUI Data Release 6.0 2023-01},
  howpublished = {https://doi.org/10.24414/z818-4163},
  month        = {January},
  year         = {2023},
  note         = {Published by Royal Observatory of Belgium (ROB)}
}

@ARTICLE{2014ApJ...789..118K,
       author = {{Krishna Prasad}, S. and {Banerjee}, D. and {Van Doorsselaere}, T.},
        title = "{Frequency-dependent Damping in Propagating Slow Magneto-acoustic Waves}",
      journal = {\apj},
         year = 2014,
        month = jul,
       volume = {789},
       number = {2},
          eid = {118},
        pages = {118},
          doi = {10.1088/0004-637X/789/2/118},
archivePrefix = {arXiv},
       eprint = {1406.3565},
 primaryClass = {astro-ph.SR},
       adsurl = {https://ui.adsabs.harvard.edu/abs/2014ApJ...789..118K}
}

@ARTICLE{2019FrASS...6...57S,
       author = {{Krishna Prasad}, S. and {Jess}, David B. and {Van Doorsselaere}, Tom},
        title = "{The temperature-dependent damping of propagating slow magnetoacoustic waves}",
      journal = {Frontiers in Astronomy and Space Sciences},
         year = 2019,
        month = aug,
       volume = {6},
          eid = {57},
        pages = {57},
          doi = {10.3389/fspas.2019.00057},
archivePrefix = {arXiv},
       eprint = {1908.00384},
 primaryClass = {astro-ph.SR},
       adsurl = {https://ui.adsabs.harvard.edu/abs/2019FrASS...6...57S}
}

@ARTICLE{2011A&A...528L...4K,
       author = {{Krishna Prasad}, S. and {Banerjee}, D. and {Gupta}, G.~R.},
        title = "{Propagating intensity disturbances in polar corona as seen from AIA/SDO}",
      journal = {\aap},
         year = 2011,
        month = apr,
       volume = {528},
          eid = {L4},
        pages = {L4},
          doi = {10.1051/0004-6361/201016405},
archivePrefix = {arXiv},
       eprint = {1102.2979},
 primaryClass = {astro-ph.SR},
       adsurl = {https://ui.adsabs.harvard.edu/abs/2011A&A...528L...4K}
}

@ARTICLE{2019A&A...628A.133K,
       author = {{Kolotkov}, D.~Y. and {Nakariakov}, V.~M. and {Zavershinskii}, D.~I.},
        title = "{Damping of slow magnetoacoustic oscillations by the misbalance between heating and cooling processes in the solar corona}",
      journal = {\aap},
         year = 2019,
        month = aug,
       volume = {628},
          eid = {A133},
        pages = {A133},
          doi = {10.1051/0004-6361/201936072},
archivePrefix = {arXiv},
       eprint = {1907.07051},
 primaryClass = {astro-ph.SR},
       adsurl = {https://ui.adsabs.harvard.edu/abs/2019A&A...628A.133K}
}

@ARTICLE{2012SoPh..275...17L,
       author = {{Lemen}, James R. and {Title}, Alan M. and {Akin}, David J. and {Boerner}, Paul F. and {Chou}, Catherine and {Drake}, Jerry F. and {Duncan}, Dexter W. and {Edwards}, Christopher G. and {Friedlaender}, Frank M. and {Heyman}, Gary F. and {Hurlburt}, Neal E. and {Katz}, Noah L. and {Kushner}, Gary D. and {Levay}, Michael and {Lindgren}, Russell W. and {Mathur}, Dnyanesh P. and {McFeaters}, Edward L. and {Mitchell}, Sarah and {Rehse}, Roger A. and {Schrijver}, Carolus J. and {Springer}, Larry A. and {Stern}, Robert A. and {Tarbell}, Theodore D. and {Wuelser}, Jean-Pierre and {Wolfson}, C. Jacob and {Yanari}, Carl and {Bookbinder}, Jay A. and {Cheimets}, Peter N. and {Caldwell}, David and {Deluca}, Edward E. and {Gates}, Richard and {Golub}, Leon and {Park}, Sang and {Podgorski}, William A. and {Bush}, Rock I. and {Scherrer}, Philip H. and {Gummin}, Mark A. and {Smith}, Peter and {Auker}, Gary and {Jerram}, Paul and {Pool}, Peter and {Soufli}, Regina and {Windt}, David L. and {Beardsley}, Sarah and {Clapp}, Matthew and {Lang}, James and {Waltham}, Nicholas},
        title = "{The Atmospheric Imaging Assembly (AIA) on the Solar Dynamics Observatory (SDO)}",
      journal = {\solphys},
         year = 2012,
        month = jan,
       volume = {275},
       number = {1-2},
        pages = {17-40},
          doi = {10.1007/s11207-011-9776-8},
       adsurl = {https://ui.adsabs.harvard.edu/abs/2012SoPh..275...17L}
}

@ARTICLE{2011ApJ...734...81M,
       author = {{Marsh}, M.~S. and {De Moortel}, I. and {Walsh}, R.~W.},
        title = "{Observed Damping of the Slow Magnetoacoustic Mode}",
      journal = {\apj},
         year = 2011,
        month = jun,
       volume = {734},
       number = {2},
          eid = {81},
        pages = {81},
          doi = {10.1088/0004-637X/734/2/81},
archivePrefix = {arXiv},
       eprint = {1104.1100},
 primaryClass = {astro-ph.SR},
       adsurl = {https://ui.adsabs.harvard.edu/abs/2011ApJ...734...81M}
}

@ARTICLE{2024MNRAS.527.5302M,
       author = {{Meadowcroft}, Rebecca L. and {Zhong}, Sihui and {Kolotkov}, Dmitrii Y. and {Nakariakov}, Valery M.},
        title = "{Observation of a propagating slow magnetoacoustic wave in a coronal plasma fan with SDO/AIA and SolO/EUI}",
      journal = {\mnras},
         year = 2024,
        month = jan,
       volume = {527},
       number = {3},
        pages = {5302-5310},
          doi = {10.1093/mnras/stad3506},
       adsurl = {https://ui.adsabs.harvard.edu/abs/2024MNRAS.527.5302M}
}

@ARTICLE{2025MNRAS.536.3192M,
       author = {{Meadowcroft}, Rebecca L. and {Nakariakov}, Valery M.},
        title = "{Fine structuring of slow magnetoacoustic wave periods in a solar coronal fan}",
      journal = {\mnras},
         year = 2025,
        month = feb,
       volume = {536},
       number = {4},
        pages = {3192-3199},
          doi = {10.1093/mnras/stae2739},
       adsurl = {https://ui.adsabs.harvard.edu/abs/2025MNRAS.536.3192M}
}

@ARTICLE{2017ApJ...849...62N,
       author = {{Nakariakov}, V.~M. and {Afanasyev}, A.~N. and {Kumar}, S. and {Moon}, Y.-J.},
        title = "{Effect of Local Thermal Equilibrium Misbalance on Long-wavelength Slow Magnetoacoustic Waves}",
      journal = {\apj},
         year = 2017,
        month = nov,
       volume = {849},
       number = {1},
          eid = {62},
        pages = {62},
          doi = {10.3847/1538-4357/aa8ea3},
       adsurl = {https://ui.adsabs.harvard.edu/abs/2017ApJ...849...62N}
}

@ARTICLE{2019ApJ...874L...1N,
       author = {{Nakariakov}, V.~M. and {Kosak}, M.~K. and {Kolotkov}, D.~Y. and {Anfinogentov}, S.~A. and {Kumar}, P. and {Moon}, Y.-J.},
        title = "{Properties of Slow Magnetoacoustic Oscillations of Solar Coronal Loops by Multi-instrumental Observations}",
      journal = {\apjl},
         year = 2019,
        month = mar,
       volume = {874},
       number = {1},
          eid = {L1},
        pages = {L1},
          doi = {10.3847/2041-8213/ab0c9f},
       adsurl = {https://ui.adsabs.harvard.edu/abs/2019ApJ...874L...1N}
}

@ARTICLE{2000A&A...362.1151N,
       author = {{Nakariakov}, V.~M. and {Verwichte}, E. and {Berghmans}, D. and {Robbrecht}, E.},
        title = "{Slow magnetoacoustic waves in coronal loops}",
      journal = {\aap},
         year = 2000,
        month = oct,
       volume = {362},
        pages = {1151-1157},
       adsurl = {https://ui.adsabs.harvard.edu/abs/2000A&A...362.1151N}
}

@ARTICLE{2024RvMPP...8...19N,
       author = {{Nakariakov}, Valery M. and {Zhong}, Sihui and {Kolotkov}, Dmitrii Y. and {Meadowcroft}, Rebecca L. and {Zhong}, Yu and {Yuan}, Ding},
        title = "{Diagnostics of the solar coronal plasmas by magnetohydrodynamic waves: magnetohydrodynamic seismology}",
      journal = {Reviews of Modern Plasma Physics},
         year = 2024,
        month = apr,
       volume = {8},
       number = {1},
          eid = {19},
        pages = {19},
          doi = {10.1007/s41614-024-00160-9},
archivePrefix = {arXiv},
       eprint = {2404.00105},
 primaryClass = {astro-ph.SR},
       adsurl = {https://ui.adsabs.harvard.edu/abs/2024RvMPP...8...19N}
}

@ARTICLE{2021SoPh..296...20P,
       author = {{Prasad}, Abhinav and {Srivastava}, A.~K. and {Wang}, T.~J.},
        title = "{Role of Compressive Viscosity and Thermal Conductivity on the Damping of Slow Waves in Coronal Loops with and Without Heating-Cooling Imbalance}",
      journal = {\solphys},
         year = 2021,
        month = jan,
       volume = {296},
       number = {1},
          eid = {20},
        pages = {20},
          doi = {10.1007/s11207-021-01764-x},
archivePrefix = {arXiv},
       eprint = {2011.14519},
 primaryClass = {astro-ph.SR},
       adsurl = {https://ui.adsabs.harvard.edu/abs/2021SoPh..296...20P}
}

@ARTICLE{2019SoPh..294..174R,
       author = {{Rachmeler}, Laurel A. and {Winebarger}, Amy R. and {Savage}, Sabrina L. and {Golub}, Leon and {Kobayashi}, Ken and {Vigil}, Genevieve D. and {Brooks}, David H. and {Cirtain}, Jonathan W. and {De Pontieu}, Bart and {McKenzie}, David E. and {Morton}, Richard J. and {Peter}, Hardi and {Testa}, Paola and {Tiwari}, Sanjiv K. and {Walsh}, Robert W. and {Warren}, Harry P. and {Alexander}, Caroline and {Ansell}, Darren and {Beabout}, Brent L. and {Beabout}, Dyana L. and {Bethge}, Christian W. and {Champey}, Patrick R. and {Cheimets}, Peter N. and {Cooper}, Mark A. and {Creel}, Helen K. and {Gates}, Richard and {Gomez}, Carlos and {Guillory}, Anthony and {Haight}, Harlan and {Hogue}, William D. and {Holloway}, Todd and {Hyde}, David W. and {Kenyon}, Richard and {Marshall}, Joseph N. and {McCracken}, Jeff E. and {McCracken}, Kenneth and {Mitchell}, Karen O. and {Ordway}, Mark and {Owen}, Tim and {Ranganathan}, Jagan and {Robertson}, Bryan A. and {Payne}, M. Janie and {Podgorski}, William and {Pryor}, Jonathan and {Samra}, Jenna and {Sloan}, Mark D. and {Soohoo}, Howard A. and {Steele}, D. Brandon and {Thompson}, Furman V. and {Thornton}, Gary S. and {Watkinson}, Benjamin and {Windt}, David},
        title = "{The High-Resolution Coronal Imager, Flight 2.1}",
      journal = {\solphys},
         year = 2019,
        month = dec,
       volume = {294},
       number = {12},
          eid = {174},
        pages = {174},
          doi = {10.1007/s11207-019-1551-2},
archivePrefix = {arXiv},
       eprint = {1909.05942},
 primaryClass = {astro-ph.SR},
       adsurl = {https://ui.adsabs.harvard.edu/abs/2019SoPh..294..174R}
}

@ARTICLE{2001A&A...370..591R,
       author = {{Robbrecht}, E. and {Verwichte}, E. and {Berghmans}, D. and {Hochedez}, J.~F. and {Poedts}, S. and {Nakariakov}, V.~M.},
        title = "{Slow magnetoacoustic waves in coronal loops: EIT and TRACE}",
      journal = {\aap},
         year = 2001,
        month = may,
       volume = {370},
        pages = {591-601},
          doi = {10.1051/0004-6361:20010226},
       adsurl = {https://ui.adsabs.harvard.edu/abs/2001A&A...370..591R}
}

@ARTICLE{2020A&A...642A...8R,
       author = {{Rochus}, P. and {Auch{\`e}re}, F. and {Berghmans}, D. and {Harra}, L. and {Schmutz}, W. and {Sch{\"u}hle}, U. and {Addison}, P. and {Appourchaux}, T. and {Aznar Cuadrado}, R. and {Baker}, D. and {Barbay}, J. and {Bates}, D. and {BenMoussa}, A. and {Bergmann}, M. and {Beurthe}, C. and {Borgo}, B. and {Bonte}, K. and {Bouzit}, M. and {Bradley}, L. and {B{\"u}chel}, V. and {Buchlin}, E. and {B{\"u}chner}, J. and {Cab{\'e}}, F. and {Cadiergues}, L. and {Chaigneau}, M. and {Chares}, B. and {Choque Cortez}, C. and {Coker}, P. and {Condamin}, M. and {Coumar}, S. and {Curdt}, W. and {Cutler}, J. and {Davies}, D. and {Davison}, G. and {Defise}, J. -M. and {Del Zanna}, G. and {Delmotte}, F. and {Delouille}, V. and {Dolla}, L. and {Dumesnil}, C. and {D{\"u}rig}, F. and {Enge}, R. and {Fran{\c{c}}ois}, S. and {Fourmond}, J. -J. and {Gillis}, J. -M. and {Giordanengo}, B. and {Gissot}, S. and {Green}, L.~M. and {Guerreiro}, N. and {Guilbaud}, A. and {Gyo}, M. and {Haberreiter}, M. and {Hafiz}, A. and {Hailey}, M. and {Halain}, J. -P. and {Hansotte}, J. and {Hecquet}, C. and {Heerlein}, K. and {Hellin}, M. -L. and {Hemsley}, S. and {Hermans}, A. and {Hervier}, V. and {Hochedez}, J. -F. and {Houbrechts}, Y. and {Ihsan}, K. and {Jacques}, L. and {J{\'e}r{\^o}me}, A. and {Jones}, J. and {Kahle}, M. and {Kennedy}, T. and {Klaproth}, M. and {Kolleck}, M. and {Koller}, S. and {Kotsialos}, E. and {Kraaikamp}, E. and {Langer}, P. and {Lawrenson}, A. and {Le Clech'}, J. -C. and {Lenaerts}, C. and {Liebecq}, S. and {Linder}, D. and {Long}, D.~M. and {Mampaey}, B. and {Markiewicz-Innes}, D. and {Marquet}, B. and {Marsch}, E. and {Matthews}, S. and {Mazy}, E. and {Mazzoli}, A. and {Meining}, S. and {Meltchakov}, E. and {Mercier}, R. and {Meyer}, S. and {Monecke}, M. and {Monfort}, F. and {Morinaud}, G. and {Moron}, F. and {Mountney}, L. and {M{\"u}ller}, R. and {Nicula}, B. and {Parenti}, S. and {Peter}, H. and {Pfiffner}, D. and {Philippon}, A. and {Phillips}, I. and {Plesseria}, J. -Y. and {Pylyser}, E. and {Rabecki}, F. and {Ravet-Krill}, M. -F. and {Rebellato}, J. and {Renotte}, E. and {Rodriguez}, L. and {Roose}, S. and {Rosin}, J. and {Rossi}, L. and {Roth}, P. and {Rouesnel}, F. and {Roulliay}, M. and {Rousseau}, A. and {Ruane}, K. and {Scanlan}, J. and {Schlatter}, P. and {Seaton}, D.~B. and {Silliman}, K. and {Smit}, S. and {Smith}, P.~J. and {Solanki}, S.~K. and {Spescha}, M. and {Spencer}, A. and {Stegen}, K. and {Stockman}, Y. and {Szwec}, N. and {Tamiatto}, C. and {Tandy}, J. and {Teriaca}, L. and {Theobald}, C. and {Tychon}, I. and {van Driel-Gesztelyi}, L. and {Verbeeck}, C. and {Vial}, J. -C. and {Werner}, S. and {West}, M.~J. and {Westwood}, D. and {Wiegelmann}, T. and {Willis}, G. and {Winter}, B. and {Zerr}, A. and {Zhang}, X. and {Zhukov}, A.~N.},
        title = "{The Solar Orbiter EUI instrument: The Extreme Ultraviolet Imager}",
      journal = {\aap},
         year = 2020,
        month = oct,
       volume = {642},
          eid = {A8},
        pages = {A8},
          doi = {10.1051/0004-6361/201936663},
       adsurl = {https://ui.adsabs.harvard.edu/abs/2020A&A...642A...8R}
}

@ARTICLE{2022SoPh..297...20S,
       author = {{Shen}, Yuandeng and {Zhou}, Xinping and {Duan}, Yadan and {Tang}, Zehao and {Zhou}, Chengrui and {Tan}, Song},
        title = "{Coronal Quasi-periodic Fast-mode Propagating Wave Trains}",
      journal = {\solphys},
         year = 2022,
        month = feb,
       volume = {297},
       number = {2},
          eid = {20},
        pages = {20},
          doi = {10.1007/s11207-022-01953-2},
archivePrefix = {arXiv},
       eprint = {2112.14959},
 primaryClass = {astro-ph.SR},
       adsurl = {https://ui.adsabs.harvard.edu/abs/2022SoPh..297...20S}
}

@ARTICLE{2025ApJ...987...57T,
       author = {{Tripathy}, Suraj K. and {Krishna Prasad}, S. and {Banerjee}, D.},
        title = "{Properties of Slow Magnetoacoustic Waves Observed Simultaneously Using Hi-C 2.1 and AIA}",
      journal = {\apj},
         year = 2025,
        month = jul,
       volume = {987},
       number = {1},
          eid = {57},
        pages = {57},
          doi = {10.3847/1538-4357/ade3c8},
archivePrefix = {arXiv},
       eprint = {2506.06126},
 primaryClass = {astro-ph.SR},
       adsurl = {https://ui.adsabs.harvard.edu/abs/2025ApJ...987...57T}
}

@ARTICLE{2013ApJ...778...26U,
       author = {{Uritsky}, Vadim M. and {Davila}, Joseph M. and {Viall}, Nicholeen M. and {Ofman}, Leon},
        title = "{Measuring Temperature-dependent Propagating Disturbances in Coronal Fan Loops Using Multiple SDO/AIA Channels and the Surfing Transform Technique}",
      journal = {\apj},
         year = 2013,
        month = nov,
       volume = {778},
       number = {1},
          eid = {26},
        pages = {26},
          doi = {10.1088/0004-637X/778/1/26},
archivePrefix = {arXiv},
       eprint = {1308.6195},
 primaryClass = {astro-ph.SR},
       adsurl = {https://ui.adsabs.harvard.edu/abs/2013ApJ...778...26U}
}

@ARTICLE{2024A&A...683A.109V,
       author = {{Van Doorsselaere}, T. and {Krishna Prasad}, S. and {Pant}, V. and {Banerjee}, D. and {Hood}, A.},
        title = "{Multithermal apparent damping of slow waves due to strands with a Gaussian temperature distribution}",
      journal = {\aap},
         year = 2024,
        month = mar,
       volume = {683},
          eid = {A109},
        pages = {A109},
          doi = {10.1051/0004-6361/202347579},
archivePrefix = {arXiv},
       eprint = {2401.09803},
 primaryClass = {astro-ph.SR},
       adsurl = {https://ui.adsabs.harvard.edu/abs/2024A&A...683A.109V}
}

@ARTICLE{2005A&A...437L..47V,
       author = {{Voitenko}, Y. and {Andries}, J. and {Copil}, P.~D. and {Goossens}, M.},
        title = "{Damping of phase-mixed slow magneto-acoustic waves: Real or apparent?}",
      journal = {\aap},
         year = 2005,
        month = jul,
       volume = {437},
       number = {3},
        pages = {L47-L50},
          doi = {10.1051/0004-6361:200500134},
       adsurl = {https://ui.adsabs.harvard.edu/abs/2005A&A...437L..47V}
}

@ARTICLE{2016GMS...216..395W,
       author = {{Wang}, T.~J.},
        title = "{Waves in Solar Coronal Loops}",
      journal = {Geophysical Monograph Series},
         year = 2016,
        month = feb,
       volume = {216},
        pages = {395-418},
          doi = {10.1002/9781119055006.ch23},
archivePrefix = {arXiv},
       eprint = {1803.11329},
 primaryClass = {astro-ph.SR},
       adsurl = {https://ui.adsabs.harvard.edu/abs/2016GMS...216..395W}
}

@ARTICLE{2021SSRv..217...34W,
       author = {{Wang}, Tongjiang and {Ofman}, Leon and {Yuan}, Ding and {Reale}, Fabio and {Kolotkov}, Dmitrii Y. and {Srivastava}, Abhishek K.},
        title = "{Slow-Mode Magnetoacoustic Waves in Coronal Loops}",
      journal = {\ssr},
         year = 2021,
        month = mar,
       volume = {217},
       number = {2},
          eid = {34},
        pages = {34},
          doi = {10.1007/s11214-021-00811-0},
archivePrefix = {arXiv},
       eprint = {2102.11376},
 primaryClass = {astro-ph.SR},
       adsurl = {https://ui.adsabs.harvard.edu/abs/2021SSRv..217...34W}
}

@ARTICLE{2019PhPl...26h2113Z,
       author = {{Zavershinskii}, D.~I. and {Kolotkov}, D.~Y. and {Nakariakov}, V.~M. and {Molevich}, N.~E. and {Ryashchikov}, D.~S.},
        title = "{Formation of quasi-periodic slow magnetoacoustic wave trains by the heating/cooling misbalance}",
      journal = {Physics of Plasmas},
         year = 2019,
        month = aug,
       volume = {26},
       number = {8},
          eid = {082113},
        pages = {082113},
          doi = {10.1063/1.5115224},
archivePrefix = {arXiv},
       eprint = {1907.08168},
 primaryClass = {astro-ph.SR},
       adsurl = {https://ui.adsabs.harvard.edu/abs/2019PhPl...26h2113Z}
}

@ARTICLE{2021SoPh..296...96Z,
       author = {{Zavershinskii}, D. and {Kolotkov}, D. and {Riashchikov}, D. and {Molevich}, N.},
        title = "{Mixed Properties of Slow Magnetoacoustic and Entropy Waves in a Plasma with Heating/Cooling Misbalance}",
      journal = {\solphys},
         year = 2021,
        month = jun,
       volume = {296},
       number = {6},
          eid = {96},
        pages = {96},
          doi = {10.1007/s11207-021-01841-1},
archivePrefix = {arXiv},
       eprint = {2104.12652},
 primaryClass = {astro-ph.SR},
       adsurl = {https://ui.adsabs.harvard.edu/abs/2021SoPh..296...96Z}
}

@ARTICLE{sunpy_community2020,
  doi = {10.3847/1538-4357/ab4f7a},
  url = {https://iopscience.iop.org/article/10.3847/1538-4357/ab4f7a},
  author = {{The SunPy Community} and Barnes, Will T. and Bobra, Monica G. and Christe, Steven D. and Freij, Nabil and Hayes, Laura A. and Ireland, Jack and Mumford, Stuart and Perez-Suarez, David and Ryan, Daniel F. and Shih, Albert Y. and Chanda, Prateek and Glogowski, Kolja and Hewett, Russell and Hughitt, V. Keith and Hill, Andrew and Hiware, Kaustubh and Inglis, Andrew and Kirk, Michael S. F. and Konge, Sudarshan and Mason, James Paul and Maloney, Shane Anthony and Murray, Sophie A. and Panda, Asish and Park, Jongyeob and Pereira, Tiago M. D. and Reardon, Kevin and Savage, Sabrina and Sipőcz, Brigitta M. and Stansby, David and Jain, Yash and Taylor, Garrison and Yadav, Tannmay and Rajul and Dang, Trung Kien},
  title = {The SunPy Project: Open Source Development and Status of the Version 1.0 Core Package},
  journal = {The Astrophysical Journal},
  volume = {890},
  issue = {1},
  pages = {68-},
  publisher = {American Astronomical Society},
  year = {2020}
}

\end{document}